\documentclass[onecolumn,notitlepage]{revtex4-1}
\usepackage{amsfonts}
\usepackage{amsmath}
\usepackage{amssymb}
\usepackage{ulem}
\usepackage{extarrows}
\usepackage{multirow}
\usepackage{float}

\usepackage{inputenc}
\usepackage{xspace}
\usepackage{url}

\usepackage{inputenc}
\usepackage{xspace}
\usepackage{url}

\usepackage{amsmath,amssymb,graphicx,esint} 

\usepackage{psfrag}
\usepackage{subfigure}
\usepackage{tabularx}
\usepackage{footmisc}

\newcommand\cc[1]{#1^{^{\kern-6pt \circ}}\kern2pt}

\renewcommand{\a}{\alpha}

\newcommand{\m}{\mu}
\newcommand{\n}{\nu}

\def\be{\begin{equation}}
\def\ee{\end{equation}}
\def\bea{\begin{eqnarray}}
\def\eea{\end{eqnarray}}
\def\ba{\begin{array}}
\def\ea{\end{array}}
\def\bi{\begin{itemize}}
\def\ei{\end{itemize}}

\newcommand{\beq}{\begin{equation}}
\newcommand{\eeq}{\end{equation}}
\newcommand{\beqn}{\begin{eqnarray}}
\newcommand{\eeqn}{\end{eqnarray}}
\newcommand{\bga}{\begin{align}}

\def\dalemb#1#2{{\vbox{\hrule height .#2pt
\hbox{\vrule width.#2pt height#1pt \kern#1pt
\vrule width.#2pt}
\hrule height.#2pt}}}

\newcommand{\auth}{Institute of Theoretical Physics, Department of Physics\\ Aristotle University of Thessaloniki,  54124 Thessaloniki, Greece.}
\newcommand{\astro}{Section of Astrophysics, Astronomy and Mechanics\\Department of Physics Aristotle University of Thessaloniki, Thessaloniki 54124, Greece.}

\begin{document}
\title{\bf Torsion/nonmetricity duality in $f(R)$ gravity}
\author{Damianos Iosifidis}
\email{diosifid@auth.gr}
\author{Anastasios C. Petkou}
\email{petkou@physics.auth.gr}
\affiliation{\auth}
\author{Christos G. Tsagas}
\email{tsagas@astro.auth.gr}
\affiliation{\astro}



\begin{abstract}
Torsion and nonmetricity are inherent  ingredients in modifications of Eintein's gravity that are based on affine spacetime geometries. In the context of pure $f(R)$ gravity we discuss here, in some detail, the relatively unnoticed duality between torsion and nonmetricity. In particular we show that for $R^2$ gravity torsion and nonmetricity are related by projective transformations. Since the latter correspond simply to redefining the affine parameters of autoparallels, we conclude that torsion and nonmetricity are physically equivalent properties of spacetime. As a simple example we show that both torsion and nonmetricity can act as geometric sources of accelerated expansion in a spatially homogenous cosmological model  within $R^2$ gravity and we briefly discuss possible implications of our results.
\end{abstract}

\maketitle


\section{Introduction}
The  study of modifications of Einstein's gravity is a major theme in current gravitational physics research. The primary interest behind such searches is the quest for a purely geometrical description for the observed cosmological evolution, without the necessity to to rely into the (as yet) elusive ingredients of dark matter and dark energy. Although this is a rather optimistic target, the study of modified theories of gravity is {\it per se} an interesting intellectual exercise which might also open  alternative and unexpected windows into physics as  the celebrated example of AdS/CFT and its numerous spinoffs have recently taught us.

In Einstein's gravity spacetime is a four dimensional Riemannian manifold equipped with a Levi-Civita (i.e. metric compatible and torsionless) connection that is fully determined by the symmetric metric. Going beyond, the most economical yet fully geometrical modifications of Einstein's theory come in the form of metric-affine theories of gravity \cite{Hehl:1994ue}. In such theories the modification comes from the introduction of a non-symmetric connection (i.e. torsion) which is not necessarily compatible with the metric (i.e. nonmetricity). Although torsion and nonmetricity are inherent ingredients of any geometric description of spacetime, they were hastily dismissed from a physical theory of gravity. Indeed, a physical implication of torsion is that parallel transport along a closed path results in a translation \cite{Hehl:2007bn}. On the other hand, nonmetricity would imply that the norm of a vector changes when it is parallel transported along itself \cite{Hehl:1994ue}. These effects, if they exist, were deemed unobservable in simple gravitational systems. Nevertheless, since they are geometrical by nature, there does not appear to be any apriori deep reason that excludes them from the generic description of  gravitation as spacetime geometry. In that sense, it is left to the various forms of matter that are coupled to gravity to probe whether or not torsion or nonmetricity have a physical relevance and possibly lead to observable implication \footnote{For recent discussions regarding the holographic applications of torsion to condensed matter systems see also \cite{Leigh:2008tt,Hughes:2011hv,Hughes:2012vg,Parrikar:2014usa}. Analogous results for nonmetricity have not yet appeared in the literature. 
} see e.g. the review \cite{Shapiro:2001rz}. 

Metric-affine  gravity is often studied using the Palatini formalism  where the spacetime metric and the connection are considered as independent dynamical variables in a first-order Lagrangian formalism.
This way, the relationship between the connection and the metric emerges as a consequence of the choice of the gravitational Lagrangian, and hence the latter becomes a nice guiding principle to study modifications of Einstein's gravity. Among the Lagrangian models used in studies of modified gravity, the so-called $f(R)$ theories have attracted an enormous amount of interested in recent years (for some recent reviews see \cite{Sotiriou:2006qn,Sotiriou:2008rp,Olmo:2011uz,Heisenberg:2018vsk}). However torsion and nonmetricity in modified gravity are less well studied and this is related to the fact that their effect in $f(R)$ gravity is mimicked by a particular Brans-Dicke  model of Einstein gravity coupled to scalars \cite{Sotiriou:2009xt}.  Nevertheless, an interesting observation of \cite{Sotiriou:2009xt} was that metric gravitational theories with torsion or nonmetric theories without torsion are equivalent to the {\it same} Brans-Dicke theory. This indicates that there may be some kind of duality in the physical implications of torsion and nonmetricity. 

In order to dwell further into this idea we revisit here the role of torsion and nonmetricity in $f(R)$ pure gravity. In Section II we review the well-known fact that only vector torsion and nonmetricity are allowed in generic $f(R)$ theories, and further that they are proportional to each other. Then we study in detail the role of projective invariance for $f(R)$ theories and show that they can be used to always obtain the Levi-Civita connection as a solution of the $f(R)\neq R^2$ equations of motion. This confirms the well-known equivalence between the metric and Palatini formulations of $f(R)\neq R^2$ gravity. In Section III we  move to the study of $f(R)=R^2$ theory. In this case we show that there is always a non-zero contribution of torsion and nonmetricity to the affine connection. Moreover, we also demonstrate that projective transformations generically interchange the roles of torsion and nonmetricity. The only projective invariant quantities are those that depend on the {\it affine vector} $w_\m$. This result implies that torsion and nonmetricity are physically equivalent as far as $R^2$ gravity is concerned and we make an attempt to give a geometric interpretation of that result. In Section IV we present a simple FRW spatially homogenous model in the context of $R^2$ gravity and show that both torsion and nonmetricity can act as geometric sources of accelerating expansion. We further analyse in some detail the nonmetric expansion that we have found. We present our conclusions and outlook in Section V. 

\section{Torsion and nommetricity in $f(R)$ gravity}
Consider the generic $f(R)$ action in $d=4$
\be
\label{GenAction}
S=\frac{1}{2\kappa^2}\int d^4x\sqrt{-g}f(R)\,,\,\,\,\,R=g^{\m\n}R_{\m\n}(\Gamma)\,,
\ee
where $\kappa^2=8\pi G_4$. The independent variations of the metric and the connection yield \footnote{A summary of our notations is presented in the Appendix. }
\begin{align}
\label{mvar}
&f'R_{(\m,\n)}-\frac{1}{2}fg_{\m\n}=0\,,\\
\label{convar}
&-\nabla_\lambda(\sqrt{-g}f'g^{\m\n})+\nabla_\sigma(\sqrt{-g}f'g^{\m\sigma})\delta^{\,\,\n}_{\lambda}+2\sqrt{-g}f'\left(g^{\m\n}S_{\lambda}-g^{\m\rho}S_{\rho}+g^{\m\sigma}S_{\sigma\lambda}^{\,\,\,\,\,\,\n}\right)=0\,.
\end{align}
Taking the trace of (\ref{mvar}) one obtains
\be
\label{mvartrace}
f'(R)R-2f(R)=0\,,
\ee
where the prime denotes derivative wrt $R$.  Taking the $\n,\lambda$ trace in (\ref{convar}) and substituting it back we obtain
\be
\label{convar1}
-\nabla_\lambda(\sqrt{-g}f'g^{\m\n})+2\sqrt{-g}f'\left(g^{\m\n}S_{\lambda}-\frac{1}{3}g^{\m\rho}S_{\rho}\delta_{\lambda}^{\,\,\n}+g^{\m\sigma}S_{\sigma\lambda}^{\,\,\,\,\,\,\n}\right)=0\,.
\ee
Then, if we take the $\m,\n$ trace in (\ref{convar1})  and use the general formula
\be
\label{nmetricity}
\nabla_\m\sqrt{-g}=-\frac{1}{2}\sqrt{-g}\,Q_\lambda\,,
\ee
we find
\be
\label{QS1}
w_\m\equiv \frac{1}{4}Q_\m+\frac{4}{3}S_\m=\partial_\m\ln f'\,.
\ee
We call the above linear combination of torsion and nonmetricity the {\it affine vector}. With the above results, we can go back to (\ref{convar1}) and obtain a general formula relating torsion and nonmetricity as
\be
\label{QS2}
Q_{\lambda}^{\,\,\,\m\n}-\frac{1}{4}Q_\lambda g^{\m\n}=\frac{2}{3}g^{\m\n}S_\lambda-\frac{2}{3}S_\rho g^{\rho\m}\delta_{\lambda}^{\,\,\n}+2g^{\m\sigma}S_{\sigma\lambda}^{\,\,\,\,\,\,\nu}\,.
\ee
Notice that taking the $(\m,\n)$ trace of (\ref{QS2}) leads to an identity that leaves unrelated the vectors $S_\m$ and $Q_\m$. 
Finally, using the formula (\ref{affinegen}) in the Appendix that gives the generic decomposition of an affine connection we obtain from (\ref{QS1})
\begin{align}
\label{affineG}
\Gamma^{\lambda}_{\,\,\m\n}&=\tilde{\Gamma}^{\lambda}_{\,\,\m\n} -\left(\frac{2}{3}S_{\n}-\frac{1}{2}w_\n\right)\delta_{\m}^{\,\,\lambda}+\frac{1}{2}\left(w_\m\delta_\n^{\,\,\lambda}-g_{\m\n}g^{\lambda\sigma}w_\sigma\right) \nonumber\\
&=\tilde{\Gamma}^{\lambda}_{\,\,\m\n} +\frac{1}{8}Q_{\n}\delta_{\m}^{\,\,\lambda}+\frac{1}{2}\left(w_\m\delta_\n^{\,\,\lambda}-g_{\m\n}g^{\lambda\sigma}w_\sigma\right)\,.
\end{align}
In (\ref{affineG}) $\tilde{\Gamma}^{\lambda}_{\,\,\m\n}$ denotes the Levi-Civita connection i.e. the Christoffel symbols of the metric $g_{\m\n}$. The torsion and nonmetricity tensors associated to the connection (\ref{affineG}) are 
\be
\label{SQ}
S_{\m\n}^{\,\,\,\,\,\,\lambda}=\frac{1}{3}(S_\m\delta_{\n}^{\,\,\lambda}-S_\n\delta_{\m}^{\,\,\lambda})\,,\,\,\,\,\,\,Q_{\lambda\m\n}=\frac{1}{4}Q_\lambda g_{\m\n}\,,
\ee
namely are fully determined from the vectors $S_\m$ and $Q_\m$ respectively \footnote{Nonmetricity of the form given in (\ref{SQ}) is called Weyl nonmetricity.}, the latter being at this point totally independent quantities. 

To proceed, we consider (\ref{mvartrace}) which is an algebraic equation for the scalar curvature $R$ yielding generically zero or constant curvature metrics \cite{Ferraris:1992dx}, except in the case when $f(R)\propto R^2$ which is the focus of our work later. In those cases the affine vector vanishes $w_\m=0$, which in turn implies that the torsion and nonmetricity vectors are proportional to each other.  Hence the affine connection (\ref{affineG}) becomes
\be
\label{affineG1}
\Gamma^{\lambda}_{\,\,\m\n}=\tilde{\Gamma}^{\lambda}_{\,\,\m\n} +\frac{1}{8}Q_{\n}\delta_{\m}^{\,\,\lambda}\,.
\ee

\subsection{Projective invariance}

Next one notices that in $f(R)$ theories the connection can only be determined up to a vectorial degree of freedom. Indeed, it is not hard to show, i.e. using the definition (\ref{Riccidef}),  that under transformations  of the form
\beq
\label{projective}
\Gamma^{\lambda}_{\,\,\mu\nu}   \rightarrow \Gamma^{\lambda}_{\,\,\mu\nu}+\delta^{\;\lambda}_{\mu}\,\xi_{\nu}\,,
\eeq
with $\xi_\m$ an arbitrary vector, the symmetric part of the Ricci tensor $R_{(\m,\n)}$ and consequently the  Ricci scalar $R$ remain invariant. 
Then, the generic $f(R)$ action is invariant under (\ref{projective}) and one can always arrive at the Levi-Civita connection by choosing an appropriate gauge parameter $\xi_\m$ in (\ref{projective}) to get rid of the terms proportional to $\delta_{\m}^{\;^\lambda}$ in (\ref{affineG1}) . 

The projective transformations (\ref{projective}) are defined as those transformations of the affine connection that leave invariant the autoparallels of vectors up to reparametrizations of the affine parameter. 
Given the curve $C:x^{\mu}=x^{\mu}{(\lambda)}$ parametrized by the affine parameter $\lambda$, and its tangent vector $u^\m(\lambda)=dx^\m/d\lambda$, we define the autoparallel curves as those satisfying
\beq
\label{autopar}
u^\lambda\nabla_\lambda u^\alpha=\frac{d^{2}x^{\alpha}}{d\lambda^{2}}+\Gamma^{\alpha}_{\,\,\mu\nu}\frac{d x^{\mu}}{d\lambda}\frac{d x^{\nu}}{d\lambda}=0\,.
\eeq
If $\Gamma^{\rho}_{\,\,\mu\nu}$ is the Levi-Civita connection this is just the geodesic equation that arises as usually when we extremize the spacetime distance $\sqrt{-g_{\m\n}dx^\m dx^\n}$. Clearly, only the symmetric part of the affine connection $\Gamma^{\lambda}_{\,\,\mu\nu}$ contributes to the autoparallel equation (\ref{autopar}), yet this part generically given by  (\ref{symmN}) receives contributions both from torsion and nonmetricity. 

There is nevertheless a freedom to define the affine connection in (\ref{autopar}) such that it only corresponds to a reparametrization of the affine parameter $\lambda$. To this end consider the transformation (\ref{projective}), when the autoparallel equation becomes
\beq
\label{autopar1}
\frac{d^{2}x^{\alpha}}{d\lambda^{2}}+\Gamma^{\alpha}_{\,\,\mu\nu}\frac{d x^{\mu}}{d\lambda}\frac{d x^{\nu}}{d\lambda}=-\xi_\n \frac{d x^{\nu}}{d\lambda}\frac{dx^\a}{d\lambda}
=f(\lambda) \frac{d x^{\alpha}}{d \lambda}  \,,
\eeq
where we have set 
\beq
f(\lambda)= -  \xi_{\mu}\frac{d x^{\mu}}{d \lambda} \,.
\eeq
This implies that $\lambda$ is not an affine parameter any more. However, using the change of variables $s=s(\lambda)$ it follows
\beq
\frac{d x^{\alpha}}{d \lambda}=\frac{d x^{\alpha}}{d s}\frac{d s}{d \lambda}=\frac{d x^{\alpha}}{d \lambda} \dot{s}\,,\,\,\,\,\,
\frac{d^{2}x^{\alpha}}{d\lambda^{2}}=\frac{d^{2}x^{\alpha}}{d s^{2}}\dot{s}^{2}+\frac{d x^{\alpha}}{d \lambda} \ddot{s}\,,
\eeq
where the dot denotes differentiation with respect to $\lambda$. Plugging these into (\ref{autopar1}) we obtain
\beq
\label{autopar2}
\frac{d^{2}x^{\alpha}}{d s^{2}}+\tilde{\Gamma}^{\alpha}_{\;\;\;\;\mu\nu}\frac{d x^{\mu}}{d s}\frac{d x^{\nu}}{d s}=\frac{1}{\dot{s}^{2}}\Big(f(\lambda)\dot{s}-\ddot{s}\Big) \frac{d x^{\alpha}}{d s}\,.
\eeq
From the above  we see that if we choose $s(\lambda)$  such that 
\beq
f(\lambda)\dot{s}-\ddot{s}=0\,,
\eeq
the right hand side vanishes and $s(\lambda)$ becomes a new affine parameter of the the autorarallel equation. The  reparametrization that we need to perform is 
\beq
s(\lambda)=\int g(\lambda)d\lambda\,,\,\,\,
g(\lambda)= e^{\int f(\lambda) d \lambda}=e^{-\int \xi_{\mu}dx^{\mu}}
\eeq

The {\it projective transformations} (\ref{projective}) are the most general form of transformations that change the autoparallel curves by a reparametrization of their affine parameter. They depend on the arbitrary vector parameter $\xi_\m$. Viewing gravity as a dynamical system, it can be shown that projective transformations are actually gauge transformations in the classical sense  i.e. \cite{Julia:1998ys,Dadhich:2010xa}. This can be used to show the equivalence of the metric and the Palatini formulations in Einstein-Hilbert $f(R)\sim R$ gravity \cite{Bernal:2016lhq}. We have thus presented here the generalization of this result to $f(R)$ gravity, implicit in a number of works i.e. \cite{Sotiriou:2009xt} showing that pure $f(R)$ theories of gravity lead to the Levi-Civita connection, up to projective  transformations.  


\section{Torsion and nonmetricity in $R^2$ gravity}


As we have discussed above, pure $f(R)$ gravity admits generically solutions with constant scalar curvature $R$, which are not suitable for the description of cosmological evolution. However, in the special case when $f(R)\propto R^2$ we cannot use (\ref{mvartrace}) to fix the scalar curvature and this allows for the possibility that torsion and nonmetricity drive nontrivial cosmological solutions. 
Consider the action
\beq
\label{R2action}
S=\frac{1}{2\kappa^2}\int d^{4}x\sqrt{-g}\alpha R^{2}\,,
\eeq 
where $\alpha$ is a parameter with dimensions of inverse mass squared. The metric variation gives now 
\beq
\label{mvarR2}
2R \left( R_{(\mu,\nu)}-\frac{R}{4}g_{\mu\nu}  \right)=0\,,
\eeq
which implies that either $R=0$ or 
\beq
\label{R4eq}
R_{(\mu,\nu)}-\frac{R}{4}g_{\mu\nu} =0
\eeq
Therefore, although Ricci flat metrics solve the e.o.m. coming from (\ref{R2action}), there exist also solutions that satisfy (\ref{R4eq}). The trace of the latter equation vanishes identically and hence it does not impose an algebraic constraint on the scalar curvature, but rather gives by virtue of (\ref{QS1})
\be
\label{R2curv}
\partial_\m\ln R=w_\m\,.
\ee
This can be formally integrated to yield
\be
\label{R2curvgen}
R=R_0\,e^{\,\int\!w_\m dx^\m}\,,
\ee
with $R_0$ some constant. 

Nevertheless, as before we should be able to use  projective transformations to further restrict torsion and nonmetricity and through them the affine connection. To begin with, we notice that under (\ref{projective}) and by virtue of the first one in (\ref{SQ}) the vector torsion transforms as
\be
\label{StoS}
S_\m \rightarrow S_\m\Bigl|_{new}=S_\m\Bigl|_{old}-\frac{3}{2}\xi_\m\,.
\ee
Then, using the projective transformation of the distortion tensor (\ref{deflcont}) and (\ref{SQ}) we find the transformation of the nonmetricity vector to be
\be
\label{QtoQ}
N^\lambda_{\,\,\m\n}\rightarrow N^\lambda_{\,\,\m\n}\Bigl|_{new}=N^\lambda_{\,\,\m\n}\Bigl|_{old} +\xi_\n\delta^\lambda_{\,\m}\,\,\Rightarrow \,\,Q_\m\rightarrow Q_{\m}\Bigl|_{new}=Q_\m\Bigl|_{old}+8\xi_\m\,.
\ee
From (\ref{StoS}) and (\ref{QtoQ}) we then learn that under projective transformations 
the affine vector  $w_{\m}$ remains invariant. This is of course consistent with the fact that $w_\m$ depends on the projective invariant scalar curvature $R$. We conclude that unless we assume zero curvature solutions, metric affine $R^2$ gravity gives rise to spacetimes with non-constant Ricci curvature. This effect is totally due to the presence of torsion and nonmetricity. 

\subsection{Torsion/nonmetricity duality and its physical implication}

The above analysis shows that although the presence of $w_\m$ cannot be gauged away by projective transformations, the individual roles of torsion and nonmetricity are actually gauge dependent. For example, we could choose $\xi_\m$ such that we eliminate {\it either} torsion {\it or} nonmetricity, namely
\begin{align}
\label{Sto0}
\xi_\m=\frac{2}{3}S_{\m}\Bigl|_{old}&\Rightarrow \,\,S_{\m}\Bigl|_{new}=0\,,\,Q_{\m}\Bigl|_{new}=Q_{\m}\Bigl|_{old}+\frac{16}{3}S_{\m}\Bigl|_{old}\neq 0\,, \\
\label{Qto0}
\xi_\m=-\frac{1}{8}Q_{\m}\Bigl|_{old}&\Rightarrow \,\,Q_{\m}\Bigl|_{new}=0\,,\,S_{\m}\Bigl|_{new}=S_{\m}\Bigl|_{old}+\frac{3}{16}Q_{\m}\Bigl|_{old}\neq 0\,.
\end{align}
More intriguingly we may chose $\xi_\m$ such that we interchange torsion and nonmetricity!. Explicitly,
\be
\label{StoQ}
\xi_\m=\frac{2}{3}S_\m\Bigl|_{old}-\frac{1}{8}Q_{\m}\Bigl|_{old}\,\,\Rightarrow\,\,S_{\m}\Bigl|_{new}=\frac{3}{16}Q_{\m}\Bigl|_{old}\,,\,Q_{\m}\Bigl|_{new}=\frac{16}{3}S_{\m}\Bigl|_{old}\,.
\ee
Notice that in the latter case the transformation leaves invariant the direct product of the torsion and nonmetricity vectors
\be
\label{SQtoSQ}
S_{\m}Q^\m \rightarrow (S_{\m}Q^\m)\Bigl|_{new}=(S_{\m}Q^\m)\Bigl|_{old}\,.
\ee
Our main result is therefore that torsion and nonmetricity are gauge equivalent physical properties of metric affine $R^2$ gravity in $d=4$. The role of gauge transformations is here played by the projective transformations (\ref{projective}) of the connection.

We may try to unveil the physical implications of the torsion/nonmetricity duality by a simple geometric example. It is known (i.e. see \cite{Hehl:2007bn}) and references therein) that torsion is related to some kind of "spacetime dislocation". Consider two infinitesimal vectors $u^\m$ and $v^\m$. Let $\Delta_{v}u^\m$ by the infinitesimal change of $u^\m$ parallel transported along $v^\m$, and correspondingly $\Delta_u v^\m$  the infinitesimal change of $v^\m$ parallel transported along $u^\m$. Then, the difference 
\be
\label{T}
T^\m\equiv \Delta_vu^\m -\Delta_uv^\m=-\left(\Gamma^\m_{\,\,\rho\sigma}-\Gamma^{\m}_{\,\,\sigma\rho}\right)v^\rho u^\sigma =2S_{\rho\sigma}^{\,\,\,\,\m}u^\rho v^\sigma\,,
\ee
is the the four-dimensional analog of the usual Burgers vector in the theory of elastic dislocations. Its physical interpretation is to quantify the failure in closing of infinitesimal rectangles in a system with a dislocation effect i.e. by going around a closed loop we reach a point translated by $T^\m$ with respect to the starting point. 

On the other hand, the physical implication of nonmetricity is to alter the length of vectors that are parallel transported along spacetime trajectories, which in turn leads to an inherent inability to define the notion of constant norm vectors. For example, if the vector $u^\m$ is parallel transported along $v^\m$ then its norm $||u||=(g_{\m\n}u^\mu u^\n)^{1/2}$ changes as
\be
\label{Du}
D_v||u||\equiv v^\m\nabla_\m(g_{\rho\sigma}u^\rho u^\sigma)^{1/2} =-\frac{1}{2||u||}v^\m Q_{\m\rho\sigma}u^\rho u^\sigma\,.
\ee
If we then consider the form of torsion and nonmetricity given in (\ref{SQ}) and we further assume that $u^\m v_\m=0$, then we can find after some calculations
\be
\label{TDu}
D_v ||u|| +\frac{1}{||u||}u_\m T^\m =-\frac{1}{2}(v_\m w^\m)||u||\,.
\ee
We therefore notice that in the affine geometry described by the $R^2$ gravity there exists a physical effect on the norm of vectors that can be attributed to {\it either} torsion {\it or} nonmetricity of to both of them. Namely, the combined effect of parallel transporting the vector $u^\m$ along a direction normal to it (i.e. along the vector $v^\m$ with $v^\m u_\m=0$), together with the normalized projection of $u^\m$ along the dislocation in the closed rectangle formed by $u^\m$ and $v^\m$, is projective invariant since it depends on the affine vector $w_\m$. In that sense, systems with just spacetime dislocations are seen to be physically equivalent to systems with just Weyl nonmetricity.

\section{Torsion and nonmetricity in cosmology}

The physical equivalence of torsion and nonmetricity can also be seen in a simple cosmological model. It was noticed in \cite{Capozziello:2008kb} that torsion can act as a geometric source of accelerated cosmology in $R^2$ affine gravity. Hence, by our results above we  expect that nonmetricity can also act as a source for cosmological acceleration in a physically indistinguishable manner \footnote{A similar observation has been recently made in \cite{Jarv:2018bgs}, in the context of the so-called symmetric teleparallel general relativity (STERG).}. Consider a spatially flat FLRW universe equipped with the metric
\beq
\label{FRWmetric}
ds^{2}=-dt^{2}+a^{2}(t)(dx^{2}+dy^{2}+dz^{2})\,,
\eeq
for which the non-vanishing Christoffel symbols are
\beq
\tilde{\Gamma}^{0}_{\;\;ij}=a\dot{a}\delta_{ij}\,,\,\,\,\,
\tilde{\Gamma}^{i}_{\;\;j0}=\frac{\dot{a}}{a}\delta^{i}_{j}\,.
\eeq
Denoting as $\tilde{R}_{\m\n}$ the Riemannian parts of the Ricci tensor we then find
\beq
\tilde{R}_{00}=-3\frac{\ddot{a}}{a}\,,\,\,\,\,
\tilde{R}_{ij}=6\left[ \frac{\ddot{a}}{a}+\left(\frac{\dot{a}}{a}\right)^{2} \right]g_{ij}\,.
\eeq
Since the metric (\ref{FRWmetric}) is spatially homogenous, it can only accomodate a nonzero temporal component of the affine vector as  $w_{0}=w(t)$ and  $w_{i}=0$. Then, from (\ref{R2curvgen}) we obtain
\beq
6(\dot{H}+2H^{2})+3\dot{w}+9Hw +\frac{3}{2}w^{2}=R_0 e^{\int wdt}\,. \label{QH1}
\eeq
Furthermore, from the field equations (\ref{mvar}) we find
\beq
\frac{\ddot{a}}{a}=\frac{1}{12}R_0 e^{\int wdt}-\frac{1}{2}\dot{w}-\frac{1}{2}Hw\,,
\eeq
or equivalently
\beq
(\dot{H}+H^{2})+\frac{1}{2}\dot{w}+\frac{1}{2}Hw      =\frac{R_0}{12}e^{\int wdt}\,. \label{QH2}
\eeq
To derive (\ref{QH2}) we have used the fact that  the affine Ricci tensor can be decomposed as \footnote{This is derived using the affine connection decomposition (\ref{affinegen}) and substituting into the Ricci tensor definition (\ref{Riccidef}).} 
\begin{align}
\label{Rdecomp}
R_{\mu\nu}=&\tilde{R}_{\mu\nu}+\frac{1}{2}\left( \tilde{\nabla}_{\mu}w_{\nu}+\tilde{\nabla}_{\nu}w_{\mu}-(\tilde{\nabla}_{\alpha}w^{\alpha}) g_{\mu\nu} \right)-2\tilde{\nabla}_{\nu}w_{\mu} +\frac{1}{2}\Big( w_{\mu}w_{\nu}-(w_{\alpha}w^{\alpha})g_{\mu\nu} \Big)\,,
\end{align}
where $\tilde\nabla$ is the Riemannian covariant derivative. From (\ref{Rdecomp}) we obtain
\beq
R_{00}=\tilde{R}_{00}-\frac{3}{2}( \dot{w}+Hw)\,.
\eeq
Then, upon combining $(\ref{QH1})$ and $(\ref{QH2})$ we arrive at
\beq
\left( H+\frac{w}{2} \right)^{2}=\frac{1}{12}R_0e^{\int wdt}\,,
\eeq
which gives
\beq
\label{H}
H=H_{0}e^{\frac{1}{2}\int w dt}-\frac{w}{2}\,. 
\eeq
We have denoted as $H_{0}=\sqrt{\frac{R_0}{12}}$ a parameter that plays the role of the initial Hubble constant. This can be positive or negative. Our final result (\ref{H}) implies that the cosmological expansion in our simple FRW model (\ref{FRWmetric}) is driven {\it both} from torsion and nonmetricity through the projective invariant affine vector $w_\m$.  Therefore, at the level our very simple spatially homogenous system torsion and nonmetricity can both act as indistinguishable sources for cosmological acceleration. Similar cosmological solutions driven by torsion have been recently discussed in \cite{Iosifidis:2018diy,Kranas:2018jdc}. In particular, by integrating (\ref{H}) one finds the evolution of the scale factor
\beq
a(t)=C_{0}e^{\int H dt}\,.
\eeq
For $w=w_{0}=constant$ we arrive at
\beq
a(t)=C_{0}e^{\frac{2 H_{0}}{w_{0}}e^{\frac{w_{0} t}{2}}-\frac{w_{0}t}{2}}\,. \label{cosmsolut}
\eeq
Now, examining ($\ref{cosmsolut}$) a little further we see that assuming  $w_{0}<0$ we find for early times 
\beq
a(t)\approx a_{0}\left[ 1-H_{0}\Big( 1-\frac{|w_{0}|}{2 H_{0}}\Big) t \right]\,,
\eeq
where $a_{0}=C_{0}e^{\frac{2 H_{0}}{w_{0}}}$. On the other hand,  for late times we have 
\beq
a(t) \approx C_{0}e^{\frac{|w_{0}|}{2}t}\,.
\eeq
Namely, we have accelerated expansion due to {\it both} torsion and non-metricity. In particular, if $|w_{0}|=2H_{0}$ the universe starts in a static state $a=a_{0}$ and accelerates exponentially for late times. It is interesting to point out that from  $(\ref{H})$, a static universe solution with $H=0$, $a=a_{0}= constant$ exists as long as
\beq
w(t)=-\frac{2}{C+t}\,,
\eeq
namely when the affine vector varies inversely with time. Notice though that for late times $w\rightarrow 0$.

\subsection{A nonmetric cosmological expansion}

As a simple example we can consider a constant affine vector that receives only contributions from nonmetricity as $w=w_{0}=Q_0/4$. Then  ($\ref{H}$) can be immediately integrated to give
\beq
a(t)=a(t)=C_{0}e^{\frac{8 H_{0}}{Q_{0}}e^{\frac{Q_{0} t}{8}}-\frac{Q_{0}t}{8}}\,,
\eeq
which is  a nonmetric cosmological expansion. Furthermore, we have
\beq
\dot{a}=a \left( H_{0}e^{\frac{Q_{0} t}{8}}-\frac{Q_{0}}{8} \right)\,,
\eeq
from which we conclude that if $H_{0}>Q_{0}/8$ we have an ever expanding universe. On the other hand, for $H_{0}=Q_{0}/8$ we have $\dot{a}(t=0)=0$ and the universe starts as static and then expands. Furthermore, when $H_{0}$ and $Q_{0}$ have the same sign, there exists a time
\beq
t^{*}=\frac{8}{Q_{0}}\ln{\left( \frac{Q_{0}}{8 H_{0}} \right)}\,,
\eeq 
after which the acceleration changes sign. Again, notice that for early times we have 
\beq
a(t)\approx a_{0}\left[ 1-H_{0}\Big( 1-\frac{|Q_{0}|}{8 H_{0}}\Big) t \right]\,,
\eeq
given that $Q_{0}<0$, while for late times
\beq
a(t) \approx C_{0}e^{\frac{|Q_{0}|}{8}t}\,,
\eeq
which is a non-metric accelerated expansion.

To complete our analysis we note that for our nonmetric cosmological model above  the Ricci scalar decomposition in terms of its Riemannian and nonmetric parts is given by
\beq
\label{RQ}
R=\tilde{R}-\frac{3}{4}\tilde{\nabla}_{\mu}Q^{\mu}-\frac{3}{32}Q_{\mu}Q^{\mu} \,\,\Rightarrow  
\,\, R=\tilde{R}+\frac{3}{4}\dot{Q}+\frac{9}{4}HQ+\frac{3}{32}Q^{2}\,.
\eeq
This expression is dual to the one that has appeared in \cite{Capozziello:2008kb}, which depend only on torsion and  was
\beq
\label{RT}
R=\tilde{R}+2\dot{T}+6 HT+\frac{2}{3}T^{2}\,.
\eeq
Indeed, it can be easily seen that (\ref{RQ}) and (\ref{RT}) are mapped into each other under 
\beq
\label{TQdual}
T \leftrightarrow  \frac{3}{8}Q
\eeq
The self-duality of the Ricci scalar under the torsion/nonmetricity exchange (\ref{TQdual}) is a general result in metric affine $f(R)$ gravity. Indeed, if {\it either} only torsion {\it or} only nonmetricity are present, the Ricci scalar decompositions are
\beq
\label{RQ1}
R=\tilde{R}-2\tilde{\nabla}_{\mu}T^{\mu}-\frac{2}{3}T_{\mu}T^{\mu}\,,
\eeq
and
\beq
\label{RT1}
R=\tilde{R}-\frac{3}{4}\tilde{\nabla}_{\mu}Q^{\mu}-\frac{3}{32}Q_{\mu}Q^{\mu} \,,
\eeq
The above map to each other under (\ref{TQdual}). 

\section{Discussion}

The main message of our work is that one needs to be very careful in distinguishing torsion and nonmetricity effects in modified theories of gravity. In particular, we have shown that for pure $R^2$ gravity torsion and nonmetricity are physically equivalent, being related by projective transformations. Similar results can be deduced from the recent analysis  in a number of works e.g. \cite{Jimenez:2015fva,Barcelo:2017tes,BeltranJimenez:2017tkd,BeltranJimenez:2018vdo,deBerredo-Peixoto:2018gqr,Iosifidis:2018diy,Damos}. Nevertheless, we believe that we have added a useful ingredient by explicitly demonstrating the role of projective transformations in the torsion/nonmetricity duality. 

It is clear the physical applications of our result should involve the study of matter coupled to modified gravity. For example one can study how the torsion/nonmetricity duality is preserved or broken by particular forms of matter. It would also be interesting to present explicit observational signatures of the duality i.e. in early cosmology. In another potential application we notice that our geometrical analysis in Section IIIA bears some resemblance with the results in \cite{Lucat:2016eze} and hence they may be used in phenomenological studies of the standard model. Finally, a potentially huge area for applications of the duality is in the context of AdS/CFT, where it is expected to give rise to relations between different strongly-coupled 3$d$ systems in the boundary, perhaps on the same par with the holographic effects of electromagnetic and gravitational duality \cite{Leigh:2003ez,Leigh:2007wf}.

\acknowledgements
We would like to thank T. Koivisto and D. Roest for some useful discussions and correspondence.

\begin{appendix}

\section*{Appendix}
Denoting with $\Gamma^{\rho}_{\;\;\;\beta\mu}$ the affine connection, covariant derivatives (of e.g. a mixed tensor) are defined as 
\begin{equation}
\label{covderdef}
\nabla_{\mu}T^{\alpha}_{\,\,\beta}=\partial_{\mu}T^{\alpha}_{\,\,\beta}+\Gamma^{\alpha}_{\,\,\rho\mu}T^{\rho}_{\,\,\beta}-\Gamma^{\rho}_{\,\,\beta\mu}T^{\alpha}_{\,\,\rho}
\end{equation}
Notice the position of the various indices in  (\ref{covderdef}). The Riemann $R^{\mu}_{\,\,\nu\alpha\beta}$ and  torsion tensors $S_{\alpha\beta}^{\,\,\nu}$ are defined via the commutator of two covariant derivatives acting on a vector $u^{\mu}$ as\footnote{The index inside horizontal bars is left out from the (anti)-symmetrization. The latter are defined as $A_{[\m,\n]}=(A_{\m\n}-A_{\n\m})/2$ and $A_{\{\m,\n\}}=(A_{\m\n}+A_{\n\m})/2$.} 
\begin{align}
\label{RSdef}
&[\nabla_{\alpha} ,\nabla_{\beta}]u^{\mu}=2\nabla_{[\alpha} \nabla_{\beta]}u^{\mu}=R^{\mu}_{\,\,\nu\alpha\beta} u^{\nu}+2 S_{\alpha\beta}^{\,\,\nu}\nabla_{\nu}u^{\mu}\\
\label{RSdef}
&R^{\mu}_{\,\,\nu\alpha\beta}:=2\partial_{[\alpha}\Gamma^{\mu}_{\,\,|\nu|\beta]}+2\Gamma^{\mu}_{\,\,\rho[\alpha}\Gamma^{\rho}_{\,\,|\nu|\beta]}\,,\,\,\,\,\,
S_{\alpha\beta}^{\,\,\nu}:=\Gamma^{\nu}_{\,\,[\alpha\beta]}
\end{align}
With a generic connection the Riemann tensor is antisymmetric in its last two indices. So, we can generically define the independent contractions giving the Ricci tensor
\begin{equation}
\label{Riccidef}
R_{\nu\beta}:=  R^{\mu}_{\,\,\nu\mu\beta} = 2\partial_{[\mu}\Gamma^{\mu}_{\,\,|\nu|\beta]}+2\Gamma^{\mu}_{\,\,\rho[\mu}\Gamma^{\rho}_{\,\,|\nu|\beta]}
\end{equation}
which, is not symmetric in $\nu,\beta$ in general, and the homothetic curvature tensor
\begin{equation}
\label{homdef}
\hat{R}_{\alpha\beta}:=R^{\mu}_{\,\,\mu\alpha\beta}=2\partial_{[\alpha}\Gamma^{\mu}_{\,\,|\mu|\beta]}=\partial_{\alpha}\Gamma^{\mu}_{\,\,\mu\beta}-\partial_{\beta}\Gamma^{\mu}_{\,\,\mu\alpha}
\end{equation}
The above discussion did not require a metric. If a symmetric metric $g_{\m\n}$ is present we can also define a third independent contraction of the Riemann tensor as \begin{equation}
\label{Rdef3}
\check{R}^{\mu}_{\;\;\beta} =g^{\nu\alpha}R^{\mu}_{\,\,\nu\alpha\beta}:=2 g^{\nu\alpha}\partial_{[\alpha}\Gamma^{\mu}_{\,\,|\nu|\beta]}+2 g^{\nu\alpha} \Gamma^{\mu}_{\,\,\rho[\alpha}\Gamma^{\rho}_{\,\,|\nu|\beta]}
\end{equation}
Moreover, the Ricci scalar is still uniquely defined since
\begin{equation}
\label{Rscaldef}
\check{R}=\check{R}^{\alpha}_{\,\,\alpha}=R^{\alpha}_{\,\,\beta\mu\alpha}g^{\beta\mu}=-R^{\alpha}_{\,\,\beta\alpha\mu}g^{\beta\mu}=-R_{\beta\mu}g^{\beta\nu}=-R
\end{equation}

There are two independent contractions of the torsion tensor giving  the torsion vector $S_\m$ and the torsion pseudovector $\tilde{S}_\m$ respectively as
\begin{equation}
\label{StS}
S_{\mu} \equiv S_{\mu\lambda}^{\,\,\lambda}\,,\,\,\,\,
\tilde{S}^{\mu} \equiv \epsilon^{\mu\nu\rho\sigma}S_{\nu\rho\sigma}
\end{equation} 

The non-metricity tensor is defined as
\begin{equation}
\label{Qdef}
Q_{\alpha\mu\nu} :=-\nabla_{\alpha}g_{\mu\nu} 
\end{equation}
It depends both on the metric tensor and the connection i.e. using the definition of the covariant derivative we have 
\begin{equation}
\label{Qdef2}
Q_{\alpha\mu\nu} :=-\nabla_{\alpha}g_{\mu\nu} =-\partial_{\alpha}g_{\mu\nu}+\Gamma^{\rho}_{\,\,\mu\alpha}g_{\rho\nu}+\Gamma^{\rho}_{\,\,\nu\alpha}g_{\mu\rho}
\end{equation}
from which, the dependence on $\Gamma^{\lambda}_{\,\,\mu\nu}$ and $g_{\mu\nu}$ is apparent. Raising the last two indices we obtain
\begin{equation}
\label{Qdef3}
Q_{\rho}^{,\alpha\beta}=\nabla_{\rho}g^{\alpha\beta} 
\end{equation}
From the non-metricity tensor we can construct two independent vectors. The Weyl vector is defined as
\begin{equation}
\label{Wdef}
Q_{\alpha}:= g^{\mu\nu}Q_{\alpha\mu\nu}=Q_{\alpha\,\,\mu}^{\,\,\,\mu}
\end{equation}
A second nonmetricity vector vector can also be defined and it is given by
\begin{equation}
\label{2nmv}
\tilde{Q}_{\nu}:= g^{\mu\alpha}Q_{\alpha\mu\nu}=Q^{\mu}_{\,\,\mu\nu}=-g^{\mu\alpha}\nabla_{\alpha}g_{\mu\nu}
\end{equation}
Finally, using the results above one can decompose the general affine connection as 
\begin{equation}
\label{affinegen}
\Gamma^{\lambda}_{\,\,\mu\nu}=\tilde{\Gamma}^{\lambda}_{\,\,\mu\nu}+N^\lambda_{\,\,\m\n}\,,
\ee
where the distortion tensor $N^\lambda_{\,\,\m\n}$ is given by 
\be
\label{deflcont}
N^\lambda_{\,\,\m\n}=\underbrace{\frac{1}{2}g^{\alpha\lambda}(Q_{\mu\nu\alpha}+Q_{\nu\alpha\mu}-Q_{\alpha\mu\nu})}_\text{deflection} -\underbrace{g^{\alpha\lambda}(S_{\alpha\mu\nu}+S_{\alpha\nu\mu}-S_{\mu\nu\alpha})}_\text{contorsion}
\end{equation}
where the Levi-Civita connection is given by the usual Christoffel symbols
\begin{equation}
\label{cristoff}
\tilde{\Gamma}^{\lambda}_{\,\,\mu\nu}:=\frac{1}{2}g^{\alpha\lambda}(\partial_{\mu}g_{\nu\alpha}+\partial_{\nu}g_{\alpha\mu}-\partial_{\alpha}g_{\mu\nu})
\end{equation}
Some useful identities are
\beq
\label{NQS}
Q_{\nu\alpha\mu}=N_{(\alpha\mu)\nu}\,,\,\,\,
S_{\mu\nu\alpha}=N_{\alpha[\mu\nu]}\,,\,\,\,
N_{[\alpha\mu\nu]}=S_{[\mu\nu\alpha]}=S_{[\alpha\mu\nu]}
\eeq
where the latter refers to the totally antisymmetric part of the distortion. as can be easily checked. Notice also  only the symmetric part $N^{\lambda}_{\,\,(\mu\nu)}$ contributes to the autoparallel equation $u^\nu\nabla_\n u^\m=0$. This is equal to
\beq
\label{symmN}
N^{\lambda}_{\,\,(\mu\nu)}=g^{\alpha\lambda}\left( Q_{(\mu\nu)\alpha}-\frac{1}{2}Q_{\alpha\mu\nu}\right)-2g^{\alpha\lambda} S_{\alpha(\mu\nu)}\,.
\eeq
For vector torsion and nonmetricity as in (\ref{SQ}) this coincides with (\ref{affineG}). Notice that a completely antisymmetric torsion $(S_{\alpha\mu\nu}=S_{[\alpha\mu\nu]})$ has no effect on autoparallels. 


\end{appendix}
\bibliographystyle{ieeetr}
\bibliography{Refs}

\end{document}